\documentstyle[12pt]{article}

\newcommand{\be}{\begin{equation}}
\newcommand{\ee}{\end{equation}}
\newcommand{\Dlt}{\Delta}

\newcommand{\om}{\omega}

\newcommand{\br}{{\bf r}}
\newcommand{\bd}{{\bf d}}
\newcommand{\bE}{{\bf E}}
\newcommand{\bH}{{\bf H}}
\newcommand{\bA}{{\bf A}}
\newcommand{\bP}{{\bf P}}
\newcommand{\bS}{{\bf S}}
\newcommand{\bB}{{\bf B}}
\newcommand{\bJ}{{\bf J}}
\newcommand{\bt}{\beta}
\newcommand{\al}{\alpha}
\newcommand{\gm}{\gamma}
\newcommand{\Gm}{\Gamma}

\newcommand{\lbd}{\lambda}
\newcommand{\prt}{\partial}

\begin{document}

\begin{center}

{\Large{\bf Spin superradiance versus atomic superradiance} \\ [5mm]
V.I. Yukalov} \\ [3mm]

{\it
Institut f\"ur Theoretische Physik, \\
Freie Universit\"at Berlin, Arnimallee 14, D-14195 Berlin, Germany \\ 
and \\
Bogolubov Laboratory of Theoretical Physics, \\
Joint Institute for Nuclear Research, Dubna 141980, Russia}

\end{center}

\vskip 1cm

\begin{abstract}

A comparative analysis is given of spin superradiance and atomic 
superradiance. Their similarities and distinctions are emphasized. It 
is shown that, despite a close analogy, these phenomena are fundamentally 
different. In atomic systems, superradiance is a self-organized process, 
in which both the initial cause, being spontaneous emission, as well as the 
collectivizing mechanism of their interactions through the common radiation 
field, are of the same physical nature. Contrary to this, in actual spin 
systems with dipole interactions, the latter are the major reason for spin 
motion. Electromagnetic spin interactions through radiation are negligible 
and can never produce collective effects. The possibility of realizing 
superradiance in molecular magnets by coupling them to a resonant circuit 
is discussed.

\end{abstract}

\vskip 1cm


\vskip 1cm


\newpage

\section{Introduction}

Spin superradiance is a phenomenon that got its name because of its close 
analogy with superradiance occurring in atomic systems [1]. In the latter, 
superradiance was predicted by Dicke [2] in 1954 and nowadays it is well 
studied both theoretically and experimentally, being now a part of textbooks 
[3,4]. Spin superradiance is a relatively novel notion, being discovered 
experimentally [5,6] in 1988, and whose exhaustive microscopic theory has 
been elaborated only recently (see reviews [1,7]).

Despite a similarity between atomic and spin superradiance, these 
phenomena are not identical, and even more, there is a fundamental 
distinction between them. This principal distinction is in the mechanism 
that triggers the initial motion and in the forces which make the system 
dynamics coherent [8--12]. Because of this radical difference, spin 
superradiance is not identical to Dicke atomic superradiance. This crucial 
fact is often overlooked, and it is possible to meet in literature wrong 
speculations caused by misunderstanding of the essentially different nature
of these phenomena. A glowing example of such a confusion is comment [13]
containing several incorrect statements based on the delusion that there 
is no any difference between atomic and spin superradiance.

The aim of the present communication is to explain the difference between 
the two types of superradiance and to emphasize that what is feasible for 
atomic systems is impossible for spin systems. A correct understanding of
these points is vitally important for the proper interpolation of experiments 
with molecular magnets. Superradiance in these materials is principally 
unachievable in the same setup as for atomic systems. However, it can be 
realized in a different way, typical of spin systems, as has been 
suggested in Ref. [14].

It is also crucially important to be precise in terminology. One sometimes 
misuses the notion of superradiance applying it vagely to any coherent 
radiation. This is certainly impermissible, since without a precise fixed
terminology any scientific descriptions and discussions become pointless.
Therefore in the present paper the notion of superradiance is used in its
strictly defined and commonly accepted sense, be it atomic [2--4] or spin 
[1,7] superradiance: {\it Superradiance is a self-organized collective 
emission of electromagnetic radiation}.

This phenomenon should not be confused with spin induction or maser 
generation, as is explained in reviews [1,7]. For instance, when the spin 
motion is pushed by a strong transverse field, one has spin induction, which, 
though being a coherent process, has nothing to do with superradiance. Or, if
a coherent process occurs on the temporal scale of the dephasing time $T_2$,
this again is not related to superradiance, since it does not involve 
collective effects.

\section{Atomic superradiance}

The microscopic description of atomic superradiance is based on the 
Hamiltonian 
\be
\label{1}
\hat H = \hat H_a +\hat H_f + \hat H_{af} \; ,
\ee
containing the terms corresponding to atoms, $\hat H_a$, field $\hat H_f$,
and atom-field interactions, $\hat H_{af}$. The Hamiltonian of $N$ 
resonant atoms is
\be
\label{2}
\hat H_a =\sum_{i=1}^N \om_0 \left ( \frac{1}{2} + S_i^z \right ) \; ,
\ee
where $\om_0$ is a transition frequency, $S_i^z$ is a preudospin operator 
characterizing the population difference, and we set $\hbar\equiv 1$. In 
the field Hamiltonian
\be
\label{3}
\hat H_f = \frac{1}{8\pi} \int \left (\bE^2 +\bH^2\right )\; d\br \; ,
\ee
$\bE$ is electric field and $\bH$ is magnetic field. The latter is defined 
through the vector potential $\bA$ as $\bH=\nabla\times\bA$. For the vector
potential, the Coulomb calibration $\nabla\cdot\bA=0$ will be used. Atom-field
interactions are described by the Hamiltonian
\be
\label{4}
\hat H_{af} = -\sum_{i=1}^N \left ( \frac{1}{c}\; \bJ_i \cdot \bA_i +
\bP_i \cdot \bE_{0i}\right ) \; ,
\ee
in which $\bA_i\equiv\bA(\br_i,t)$, $\bE_{0i}\equiv\bE_0(\br_i,t)$ is an
external field, if any, the transition current is
\be
\label{5}
\bJ_i = i\om_0 \left (\bd S_i^+ -\bd^* S_i^-\right ) \; ,
\ee
where $\bd$ is a transition dipole, and the transition polarization is
\be
\label{6}
\bP_i =\bd S_i^+ + \bd^* S_i^- \; ,
\ee
with $S_i^\pm$ being the ladder operators. For the pseudospin operators,
the notation $S_i^\al\equiv S^\al(\br_i,t)$ is used. Here the electric 
dipole transitions are assumed, although practically the same consideration 
is valid for magnetic dipole transitions [15].

For the best comparison of atomic coherent effects with their spin 
counterpart, it is necessary to represent the atomic evolution equations 
in terms of the pseudospin operators. This can be done by the method of 
elimination of field variables [16], which allows for the reduction of 
the equations of motion to the Heisenberg equations for the pseudospin 
variables. Averaging these equations over the spin degrees of freedom,
we come to the equations for the {\it transverse function}
\be
\label{7}
u(\br,t) \equiv 2<S^-(\br,t)> \; ,
\ee
{\it coherence intensity}
\be
\label{8}
w(\br,t) \equiv 4< S^+(\br,t) S^-(\br+0,t)> \; ,
\ee
and the average {\it population difference}
\be
\label{9}
s(\br,t) \equiv 2<S^z(\br,t) > \; .
\ee
To write down the evolution equations in a compact form, it is convenient 
to introduce the notation for the effective force
\be
\label{10}
f(\br,t) =  f_0(\br,t) + f_{rad}(\br,t) +\xi(\br,t) \; ,
\ee
in which the first term
\be
\label{11}
f_0(\br,t) \equiv -2i\bd \cdot\bE_0(\br,t)
\ee
describes the action of an external field, the second term
\be
\label{12}
f_{rad}(\br,t) \equiv 2 k_0 <\bd \cdot \bA_{rad}(\br,t)>
\ee
corresponds to the interaction through resonant radiation with a frequency 
$\om_0$ and wave vector $k_0=\om_0/c$, and the last term
\be
\label{13}
\xi(\br,t) \equiv 2k_0\bd \cdot \left (\bA_{vac} + \bA_{dip}
\right )
\ee
is due to vacuum fluctuations and to effective transition-dipole 
interactions. The radiation force (12) can be represented as
\be
\label{14}
f_{rad}(\br,t) = -i \gm_0\rho \; \int \left [ G(\br-\br',t) u(\br',t) - \; 
\frac{\bd^2}{|\bd|^2} \; G^*(\br-\br',t) u^*(\br',t)\right ] \; d\br' \; ,
\ee
where $\rho\equiv N/V$ is atomic density, the transfer function is
$$
G(\br,t) = \frac{\exp(ik_0r)}{k_0 r} \; \Theta(ct-r) \; ,
$$
with $\Theta(\cdot)$ being the unit-step function, and where the natural
width
\be
\label{15}
\gm_0 \equiv \frac{2}{3}\; |\bd|^2 k_0^3
\ee
has entered the equations. Then the evolution equations for functions (7)
to (9) take the form
$$
\frac{\prt u}{\prt t} = - i \left (\om_0 -i\gm_2 -i\gm_2^*\right ) u + 
fs\; , \qquad
\frac{\prt w}{\prt t} = -2\left ( \gm_2+\gm_2^*\right ) w +
\left (u^* f + f^* u\right ) s \; ,
$$
\be
\label{16}
\frac{\prt s}{\prt t} = -\; \frac{1}{2}\left ( u^* f + f^* u\right ) -
\gm_1( s-\zeta) \; ,
\ee
where $\gm_2\approx\gm_0$ is a transverse homogeneous relaxation rate, 
$\gm_2^*$ is inhomogeneous broadening, $\gm_1$ is a longitudinal relaxation 
rate, and $\zeta$ is a stationary population difference for an atom. All
details of deriving Eqs. (16) can be found in review [16].

A necessary condition for the appearance of coherent effects is the 
existence of strongly correlated regions, so called {\it wave packets}, 
of a characteristic size $L_c$, such that
\be
\label{17}
k_0 L_c \ll 1 \qquad \left ( k_0 \equiv \frac{\om_0}{c} \right ) \; .
\ee
Coherence can arise only inside such a well correlated region. The 
characteristic length $L_c$ must be much larger than the mean interatomic 
distance, however it can be much smaller than the system length $L$. It is 
solely in the extreme case of the so-called concentrated system, for which 
the radiation wavelength $\lbd\gg L$, when $L_c=L$. But the appearance of
superradiance, of course, does not require that $\lbd$ be larger than the 
system length. Moreover, in standard optical systems $\lbd$ is always 
shorter than $L$. And optical superradiance does exist in such systems 
with $\lbd\ll L$, which is, actually, a well known fact expounded in 
textbooks [3,4] and noticed yet in the original paper by Dicke [2] (see
also discussion in Ref. [17]). When $\lbd\ll L$, then the sample becomes 
separated into correlated regions with $L_c\ll L$, and inside each of such 
regions coherence develops owing to condition (17). The correlation between 
different wave packets is weak, because of which the correlated regions 
radiate almost independently, thus, producing oscillations in the 
superradiant pulse. In lasers with large aperture, coherent radiation is 
always realized in the form of a bunch of filaments [18--21]. Superradiance 
with $\lbd\ll L$ can arise even in photonic bound-gap materials, where 
atomic spontaneous emission is suppressed [22]. Hence, one should not confuse 
the inequality $\lbd\gg L$, which is not required for the occurrence of
superradiance in a sample of length $L$, with the necessary condition (17),
which is compulsory for any coherence to develop in the system.

For each correlated region, the evolution equations (16) can be described 
in the single-mode picture and can be reduced to the guiding-center 
equations
\be
\label{18}
\frac{dw}{dt} = -2\gm_2 \left ( 1 +\kappa -gs\right ) w +
2\gm_3 s^2 \; ,
\ee
\be
\label{19}
\frac{ds}{dt} = - g\gm_2 w - \gm_3 s  -\gm_1 ( s - \zeta) \; ,
\ee
in which $\kappa\equiv\gm_2^*/\gm_2$, $\gm_3\approx\gm_0$ is the 
relaxation rate due to the local fluctuations caused by effective 
dipole-dipole interactions, and the  coupling parameter
\be
\label{20}
g \equiv \rho\; \frac{\gm_0}{\gm_2} \; \int 
\frac{\sin k_0(r-z)}{k_0\; r} \; d\br
\ee
describes the intensity of atomic interactions through the common radiation 
field. Under condition (17), the coupling parameter $g\sim N_c\gg 1$ is very
large. For instance, for the CO$_2$ laser with the atomic density 
$\rho\sim 10^{18}$ cm$^{-3}$ and the working wavelength $\lbd\sim 10^{-3}$ 
cm, estimating $N_c$ as $\rho\lbd^3$, we have $g\sim 10^9$. For solid state
lasers, with $\rho\sim 10^{22}$ and $\lbd\sim 10^{-4}$, one has $g\sim 
10^{10}$. Thus, condition (17) not only is compulsory for the development 
of coherence, but it is also favorable for having a large coupling parameter 
(20), which is necessary for achieving superradiance.

The development of atomic superradiance goes as follows. The system is 
prepared in a nonequilibrium state, with inverted population difference,
so that $s_0\equiv s(0)>0$. No transverse fields are applied, so that 
$w_0\equiv w(0)=0$, since superradiance implies, by definition, a 
self-organized process. Atoms start with spontaneous emission, at the
beginning being not correlated with each other. Their interactions through 
the radiated field yield the development of mutual correlations. After the 
crossover time $t_c=T_2/2gs_0$, these correlations become strong inside a 
correlated region of length $L_c$. Then the initial chaotic stage changes
to the coherent stage, and atoms of each spatial mode start radiating 
coherently. The maximum of a superradiant burst happens at the delay time
$$
t_0 = t_c + \frac{\tau_p}{2}\; \ln\left |
\frac{2}{\gm_3 t_c}\right | \; .
$$
The duration of the superradiant pulse is given by the pulse time
$\tau_p=T_2/gs_0$. Because of the very large coupling $g$, the pulse time 
can become an order shorter than $T_2$. The condition $\tau_p\ll T_2$ is
the criterion for a coherent pulse to be called superradiant. Respectively,
the intensity of radiation can be one or several orders stronger than 
that of incoherent radiation. In a concentrated system the intensity of
a superradiant burst would be proportional to $N^2$, with $N$ being the 
total number of atoms. For realistic systems, with $\lbd\ll L$, where the 
radiation comes from several spatial modes, taking account of the spatial 
dispersion leads to the radiation intensity $I\sim N^\al$, with $1<\al<2$.
The value of $\al$ depends on the sample shape. Thus for pencil-like or 
disk-like samples, one has $I\sim N^{5/3}$ or $I\sim N^{4/3}$, respectively 
[22].

It is important to stress that in an atomic system the relaxation rates 
$\gm_0$, $\gm_2$, and $\gm_3$ all are caused by the same physical origin, 
by the radiation of atoms, and so all of them are of the same order of the
natural width $\gm_0$.

\section{Spin superradiance}

The microscopic description of spin superradiance has to be based on a 
realistic spin Hamiltonian [1,7]. For the generality of the consideration 
and in order to stress the principal difference from atomic systems, the 
spin Hamiltonian can be complimented by the field Hamiltonian $\hat H_f$,
as in Eq. (3), and, respectively, by the spin interactions with 
electromagnetic field. Although electromagnetic spin interactions are 
negligible, as compared to direct dipole interactions [7], we shall keep 
them to demonstrate once again that they are not able to produce spin 
superradiance. This fact is what makes spin systems fundamentally different 
from atomic ones. To realize superradiance in a spin system, it is absolutely 
necessary to couple it to a resonant electric circuit [1,7].

The system of $N$ spins is described by the Hamiltonian
\be
\label{21}
\hat H =\sum_{i=1}^N \hat H_i + \frac{1}{2} \sum_{i\neq j}^N
\hat H_{ij} + \hat H_f \; ,
\ee
where the single-spin term includes the interaction of spins with all 
magnetic fields, the second term is due to dipole spin interactions, and 
the last term is the field Hamiltonian (3). In the single-spin Hamiltonian
\be
\label{22}
\hat H_i = -\mu_0 \bS_i \cdot \bB_i - D (S_i^z)^2 \; ,
\ee
$\mu_0\equiv\hbar\gm_S$, with $\gm_S$ being the gyromagnetic ratio of a
particle with spin $S$. In what follows, we shall set $\hbar=1$. The Zeeman
term of Eq. (22) includes the interaction of a spin $\bS_i$ with the total
magnetic field
\be
\label{23}
\bB_i = B_0{\bf e}_z + H_{res}{\bf e}_x +\bH_i \; ,
\ee
where $B_0$ is a longitudinal external field, $H_{res}$ is a feedback 
field of a resonant electric circuit coupled to the sample, and 
$\bH_i\equiv\bH(\br_i,t)$ is the radiation field. The second term in Eq. (22) 
is the energy of the single-site magnetic anisotropy, with the anisotropy 
parameter $D$. The interaction Hamiltonian in Eq. (21) is due to direct 
dipole interactions
\be
\label{24}
\hat H_{ij} = \sum_{\al\bt} D_{ij}^{\al\bt} S_i^\al S_j^\bt \; ,
\ee
with the dipolar tensor $D_{ij}^{\al\bt}$.

Writing down the Heisenberg equations of motion for spin operators 
$S_i^\al\equiv S^\al(\br_i,t)$, we shall average them over the spin 
degrees of freedom. Our aim is to obtain the evolution equations for the
{\it transverse function}
\be
\label{25}
u(\br,t) \equiv \frac{1}{S}\; < S^-(\br,t)>\; ,
\ee
{\it coherence intensity}
\be
\label{26}
w(\br,t) \equiv \frac{1}{S^2}\; < S^+(\br,t) S^-(\br+0,t)> \; ,
\ee
and the longitudinal {\it spin polarization}
\be
\label{27}
s(\br,t) \equiv \frac{1}{S}\; < S^z(\br,t) > \; .
\ee
These functions are analogous to Eqs. (7) to (9).

Let us introduce the notation for the effective force acting on a spin as
\be
\label{28}
f= f_{res} + f_{rad} + \xi \; ,
\ee
where the resonator force
$$
f_{res} = -i\mu_0 H_{res}
$$
is caused by the action of the resonator feedback field $H_{res}$, the
radiation force
$$
f_{rad} =\gm_r u
$$
is due to spin interactions through the common radiation field [7,23], 
with the radiation rate
\be
\label{29}
\gm_r \equiv \gm_0 N_c \; , \qquad 
\gm_0 \equiv \frac{2}{3}\; \mu_0^2 S k^3 \; ,
\ee
in which $N_c$ is the number of correlated spins and $\gm_0$ is the natural
radiation width; and the last term in Eq. (28) representing local spin 
fluctuations caused by dipole interactions [1,7].

The effective transverse relaxation rate, owing its origin to dipole 
spin-spin interactions, can be represented [24] as
\be
\label{30}
\gm_2(s) = (1 - s^2) \gm_2 \; ,
\ee
where
\be
\label{31}
\gm_2 \equiv \gm_2(0) = n_0 \rho\mu_0^2\sqrt{S(S+1)} \; ,
\ee
with $n_0$ being the number of nearest neighbours and $\rho\equiv N/V$ 
being the spin density. Including also the inhomogeneous broadening 
$\gm_2^*$, we have the total transverse relaxation rate
\be
\label{32}
\Gm_2 = (1-s^2)\gm_2 + \gm_2^* \; .
\ee

Also, we shall employ the notation for the Zeeman frequency
\be
\label{33}
\om_0 \equiv -\mu_0 B_0 \; ,
\ee
anisotropy frequency
\be
\label{34}
\om_D \equiv (2S-1) D \; ,
\ee
and the effective spin rotation frequency
\be
\label{35}
\om_s \equiv \om_0 - \om_D s \; .
\ee
Small frequency shifts, caused by the radiation friction and local dipole 
fluctuations, can be omitted [7]. Thus, we come to the equations
$$
\frac{du}{dt} = - i(\om_s - i\Gm_2) u + fs \; , \qquad
\frac{dw}{dt} =  -2 \Gm_2 w + \left ( u^* f + f^* u\right ) s\; ,
$$
\be
\label{36}
\frac{ds}{dt} = -\; \frac{1}{2}\left ( u^* f + f^* u\right ) -
\gm_1 (s -\zeta) \; ,
\ee
in which $\gm_1$ is the spin-lattice attenuation and $\zeta$ is the value 
of a stationary spin polarization. These equations are analogous to Eqs. 
(16). Notice, however, that the frequency of spin rotation is $\om_s$, 
given by Eq. (35), but not the Zeeman frequency (33).

In order that the interactions through electromagnetic radiation could lead 
to the appearance of noticeable correlations among spins, the correlated 
regions must exist of a typical length $L_c$, such that
\be
\label{37}
kL_c \ll 1 \qquad \left ( k \equiv \frac{\om_s}{c}\right ) \; ,
\ee
which is in complete analogy with Eq. (17). Condition (37) is compulsory 
for the appearance of any coherence through the common radiation field. 
All related discussion, given above for atomic systems, is valid as well 
for spin systems. The correlation length $L_c$ can vary between the mean
interspin distance $a$ and the linear system size $L$. Respectively, the
number of correlated spins $N_c$ varies between $\rho a^3=1$ and the total
number of spins $\rho L^3=N$.

Defining the resonator feedback field by the Kirchhoff equation, we find
[1,7] the effective spin-resonator coupling
\be
\label{38}
g \equiv \frac{\gm\overline\gm_0 \om_s}{\gm_2(\gm^2 +\Dlt^2)} \; ,
\ee
in which $\gm$ is the circuit damping, the value
$$
\overline\gm_0 \equiv \pi\eta\rho\mu_0^2 S \; ,
$$
where $\eta$ is a filling factor, is of the order of $\gm_2$ in Eq. (31), 
and $\Dlt\equiv\om-|\om_s|$ is a detuning between the circuit natural 
frequency $\om$ and the spin rotation frequency $\om_s$, given by Eq. (35). 
Taking into account that $\gm\equiv\om/2Q$, where $Q$ is a resonator quality 
factor, we have $g\sim Q$, which can be as large as $g\sim 10^3-10^5$.

The effective spin coupling through the radiation field is
\be
\label{39}
g_r \equiv \frac{\gm_r}{\gm_2} = \frac{2}{3n_0}\; \sqrt{\frac{S}{S+1}} \;
\left ( k L_c\right )^3 \; .
\ee
According to the necessary condition (37), and since $n_0\sim 10$, we get 
$g_r\ll 1$.

Solving the first of equations (36) with respect to $u$ and substituting 
this solution to the second and third of Eqs. (36), we obtain, after the
averaging procedure over fast oscillations [1,7], the guiding-center
equations
$$
\frac{dw}{dt} = - 2\gm_2\left ( 1 +\kappa -s^2 -gs -g_r s\right ) w +
2\gm_3 s^2 \; ,
$$
\be
\label{40}
\frac{ds}{dt} = -\gm_2(g+g_r) w -\gm_3 s -\gm_1 (s-\zeta) \; ,
\ee
in which $\kappa\equiv\gm_2^*/\gm_2$ and $\gm_3\leq\gm_2$ is an attenuation
owing its origin to the local spin fluctuations related to dipole interactions.
These are the evolution equations that are to be compared with Eqs. (18) 
and (19).

Because of a negligibly small value of the radiation coupling $g_r\ll 1$, 
it plays no role in spin dynamics. It is easy to check that even when there 
is no resonator coupling, that is $g\equiv 0$, the terms with $g_r\ll 1$ do
not influence the spin motion described by Eqs. (40). This implies that the
interaction of spins through the common radiation field can never produce 
spin superradiance. This conclusion is general and is valid for any type 
of spins, whether these are nuclear, electron, or molecular spins. Contrary
to the negligible radiation coupling (39), the spin-resonator coupling (38),
as is explained above, can be made very large. It is only by coupling a spin 
sample to a resonant electric circuit that one is able to realize spin
superradiance [1,7]. This drastic difference of spin systems from atomic 
ones comes from the fact that in the former the linewidth $\gm_2$ and the
radiation rate $\gm_r$ are caused by distinct physical origins, while in
atomic systems they are of the same origin. In spin systems, the radiation
rate $\gm_r\ll\gm_2$ is negligible, as compared to the dipole linewidth 
$\gm_2$. On the contrary, in atomic systems the collective radiation rate 
$g\gm_2\gg\gm_2$ is much larger than the linewidth $\gm_2$.

In a spin sample coupled to a resonant circuit, spin superradiance takes 
place in the following way. The sample is prepared in an inverted state, 
when its average magnetization is opposite to an external magnetic field.
Initial spin motion is triggered by local spin fluctuations caused by 
dipole interactions. Chaotic fluctuations last till the chaos time 
$t_c\approx\tau/gs_0$, where $\tau\equiv\gm^{-1}$ is the resonator ringing 
time and $s_0\equiv s(0)$. Then correlations in the transverse spin motion 
begin rising fast, being due to the collectivization through the resonator 
feedback field. The superradiant burst peaks at the delay time
$$
t_0 = t_c  + \frac{\tau_p}{2}\; 
\ln\left | \frac{2}{\gm_3 t_c}\right | \; .
$$
The pulse time $\tau_p=T_2/gs_0$ is much shorter than $T_2$, which is
necessary for a pulse to be treated as superradiant.

The possibility of superradiant operation by spin systems can be used in a 
number of applications [1,7], for example, by creating spin masers [14]. By 
analogy with squeezed atomic states [25], one can realize spin squeezing.
Punctuated spin superradiance can be employed for information processing 
[26].

In recent years, much attention is given to the consideration of molecular
magnets consisting of molecules whose spin can reach high values. The 
feasibility of realizing the resonator-assisted spin superradiance in such 
materials was advanced in Ref. [14]. At the same time, there have appeared 
speculations [13,27] that superradiance in molecular magnets could occur 
due only to spin interactions through the common radiation field. As is 
thoroughly explained above, this is certainly impossible. Spin superradiance 
can happen only as a resonator-assisted phenomenon [14]. There have been 
attempts of measuring possible microwave emission generated in the avalanches 
of magnetization reversal in Mn$_{12}$-acetate at low temperatures [28].
However in other highly precise experiments [29] no significant radiation 
at well-defined frequencies was detected. Actually, the low-temperature 
step-like avalanches in Mn$_{12}$ acetate are known for a long time and can 
be due to different reasons [30]. But, as the accurate theory shows, these 
avalanches cannot be related to superradiance.

Moreover, in a typical experimental setup for such a molecular magnet as 
Mn$_{12}$-acetate, the process of an avalanche is not characterized by a 
well-defined frequency. Really, let us consider the values of the parameters
for Mn$_{12}$-acetate, whose spin is $S=10$. Then $D\cong 0.6$ K, so that in
frequency units $D=0.8\times 10^{11}$ s$^{-1}$, and the anisotropy frequency 
(34) is $\om_D=1.5\times 10^{12}$ s$^{-1}$. The density is $\rho=2.2\times 
10^{20}$ cm$^{-3}$. Using $\gm_S=-1.759\times 10^7$ G$^{-1}$s$^{-1}$,
$\mu_0=-1.856\times 10^{-20}$ erg/G, and $\mu_0\gm_S=3.265\times 10^{-13}$
cm$^3$/s, we find the dipole linewidth (31) as $\gm_2=0.8\times10^{10}$
s$^{-1}$. The Zeeman frequency (33) for the field $B_0=1T=10^4$ G is 
$\om_0=1.76\times 10^{11}$ s$^{-1}$, for $B_0=2T$, one has 
$\om_0=3.52\times 10^{11}$ s$^{-1}$, and $\om_0=5.28\times 10^{11}$ 
s$^{-1}$ for $B_0=3T$. Experiments with magnetization avalanches are 
usually done in the field of $2-3T$. For such fields, the Zeeman frequency 
$\om_0\sim 10^{11}$ s$^{-1}$ is much smaller than the anisotropy frequency 
$\om_D\sim 10^{12}$ s$^{-1}$. Hence, the spin rotation frequency (35) is 
defined mainly by the term $-\om_Ds$. In the process of an avalanche, the
spin polarization $s$ changes from $s=1$ to $s=-1$. The spin frequency 
varies by the amount $\Dlt\om_s=-\om_D\Dlt s$. For the polarization change 
$\Dlt s=-2$, the frequency variation is $\Dlt\om_s=2\om_D=3\times 10^{12}$ 
s$^{-1}$. Even if $s$ changes by $\Dlt s=-0.1$, corresponding to one 
quantum transition, the spin frequency varies as 
$\Dlt\om_s=0.1\om_D=1.5\times 10^{11}$ s$^{-1}$. This means that there is 
no well-defined radiation frequency during the avalanche.

Recall that the appearance of the radiation rate (29) as such is 
preconditioned by the existence of a well-defined constant radiation 
frequency. If the latter does not exist, then the radiation rate (29) has 
no sense, as far as no coherent radiation can arise in principle.

Concluding, spin superradiance can be realized in spin systems only by 
coupling them to a resonant electric circuit. To preserve the stability of 
the spin frequency $\om_s$, one can either impose a sufficiently strong 
external magnetic field $B_0$, such that $\om_0$ be essentially larger 
than $\om_D$, or by varying $B_0$ to achieve the chirping effect, as 
described in Refs. [7,14].

\vskip 5mm

{\it Acknowledgements}. I am grateful to E.P. Yukalova for useful discussions.
The Mercator Professorship of the German Research Foundation is appreciated.

\end{document}